\documentclass[prl,aps,twocolumn,superscriptaddress]{revtex4-1}
\usepackage{amsfonts}

\usepackage{dcolumn}
\usepackage{bm}
\usepackage{tikz}
\usepackage[colorlinks=true,citecolor=blue,urlcolor=blue]{hyperref}
\usepackage{amsmath,amsfonts,amssymb,times,natbib}
\usepackage[standard]{ntheorem}

\begin{document}

\title{Quantum simulation for three-dimensional chiral topological insulator}% Force line breaks with \\

\author
{Wentao Ji$^{1,2,5\ast}$,
Lin Zhang$^{3,4\ast}$,
Mengqi Wang$^{1,2,5}$,
Long Zhang$^{3,4}$,  
Yuhang Guo$^{1,2,5}$,
Zihua Chai$^{1,2,5}$,
Xing Rong$^{1,2,5}$,
Fazhan Shi$^{1,2,5}$, 
Xiong-Jun Liu$^{3,4,6,7\dag}$,
Ya Wang$^{1,2,5\dag}$,
Jiangfeng Du $^{1,2,5\dag}$ 
\\
\normalsize{$^{1}$ Hefei National Laboratory for Physical Sciences at the Microscale and Department of Modern Physics, University of Science and Technology of China (USTC), Hefei, 230026, China.}\\
%\normalsize{Department of Modern Physics,}
%\normalsize{University of Science and Technology of China (USTC), Hefei, 230026, China.}
\normalsize{$^{2}$ CAS Key Laboratory of Microscale Magnetic Resonance, USTC, Hefei, 230026, China.}\\
\normalsize{$^{3}$ International Center for Quantum Materials, School of Physics,}
\normalsize{Peking University, Beijing 100871, China.}\\
\normalsize{$^{4}$ Collaborative Innovation Center of Quantum Matter, Beijing 100871, China.}\\
\normalsize{$^{5}$ Synergetic Innovation Center of Quantum Information and Quantum Physics,}
\normalsize{USTC, Hefei, 230026, China.}\\
\normalsize{$^{6}$ Shenzhen Institute for Quantum Science and Engineering and Department of Physics,}
\normalsize{Southern University of Science and Technology, Shenzhen 518055, China.}\\
\normalsize{$^{7}$ Beijing Academy of Quantum Information Science, Beijing 100193, China}\\
\normalsize{$^{\ast}$ These authors contributed equally to this work.}\\
\normalsize{$^\dag$ E-mail: xiongjunliu@pku.edu.cn, ywustc@ustc.edu.cn, djf@ustc.edu.cn }
}

\date{\today}
%%%%%%%%%%%%%%%%%%%%%%%%%%%%%%%%%%%%%%%%%%%%%%%%%%%%%%%%%%%%%%%%%%%
\begin{abstract}
Quantum simulation, as a state-of-art technique, provides the powerful way to explore topological quantum phases beyond natural limits. Nevertheless, a complete simulation of the bulk and surface topological physics, and their correspondence is usually hard to achieve in one single simulator. Here we build up a quantum simulator using nitrogen-vacancy center to investigate a previously-not-realized three-dimensional (3D) chiral topological insulator, and demonstrate by quantum quenches a complete study of both the bulk and surface topological physics. First, a dynamical bulk-surface correspondence in momentum space is observed, showing that the bulk topology of the 3D phase uniquely corresponds to the nontrivial quench dynamics emerging on 2D momentum hypersurfaces called band inversion surfaces (BISs), equivalent to the bulk-boundary correspondence in real space. Further, the symmetry protection of the 3D chiral phase is uncovered by measuring dynamical spin textures on BISs, which exhibit perfect (broken) topology when the chiral symmetry is preserved (broken). Finally we measure the topological charges to characterize directly the bulk topology, and identify an emergent dynamical topological transition when varying the quenches from deep to shallow regimes. This work opens a new avenue of quantum simulation towards the complete study of topological quantum phases.
\end{abstract}
%\pacs{03.65.Ud, 03.67.Mn, 42.50.Xa}
\maketitle

\emph{Introduction.}---The past over one decade has witnessed the explosive progress in the field of topological quantum phases~\cite{Hasan2010,Qi2011,Chiu2016}, with many exotic topological states having been discovered in the tabletop materials~\cite{Yan2012,Ando2013,Yan2017}. The most prominent classes of topological materials include the time-reversal invariant topological insulators~\cite{Konig2007,Hsieh2008,Xia2009,Knez2011}, quantum anomalous Hall insulators~\cite{Chang2013}, topological semimetals~\cite{Liu2014a,Lv2015,Xu2015}, and topological superconductors~\cite{Mourik2012,Sun2016,Zhang2018}. These topological phases are characterized by nontrivial topology in the bulk, and host topology- or symmetry-protected gapless boundary modes which are connected to the bulk through the bulk-boundary correspondence~\cite{Hasan2010,Qi2011,Chiu2016}. Such bulk-boundary correspondence has been the dominant mechanism for the observation of the topological states, with most topological materials having been uncovered in experiment by resolving the boundary physics~\cite{Hsieh2008,Xia2009,Lv2015,Xu2015}, while the bulk topology, however, is hard to be directly measured for solid systems.

\begin{figure*}
	\centering
	%	\fbox
	{\includegraphics[width=1.8\columnwidth]{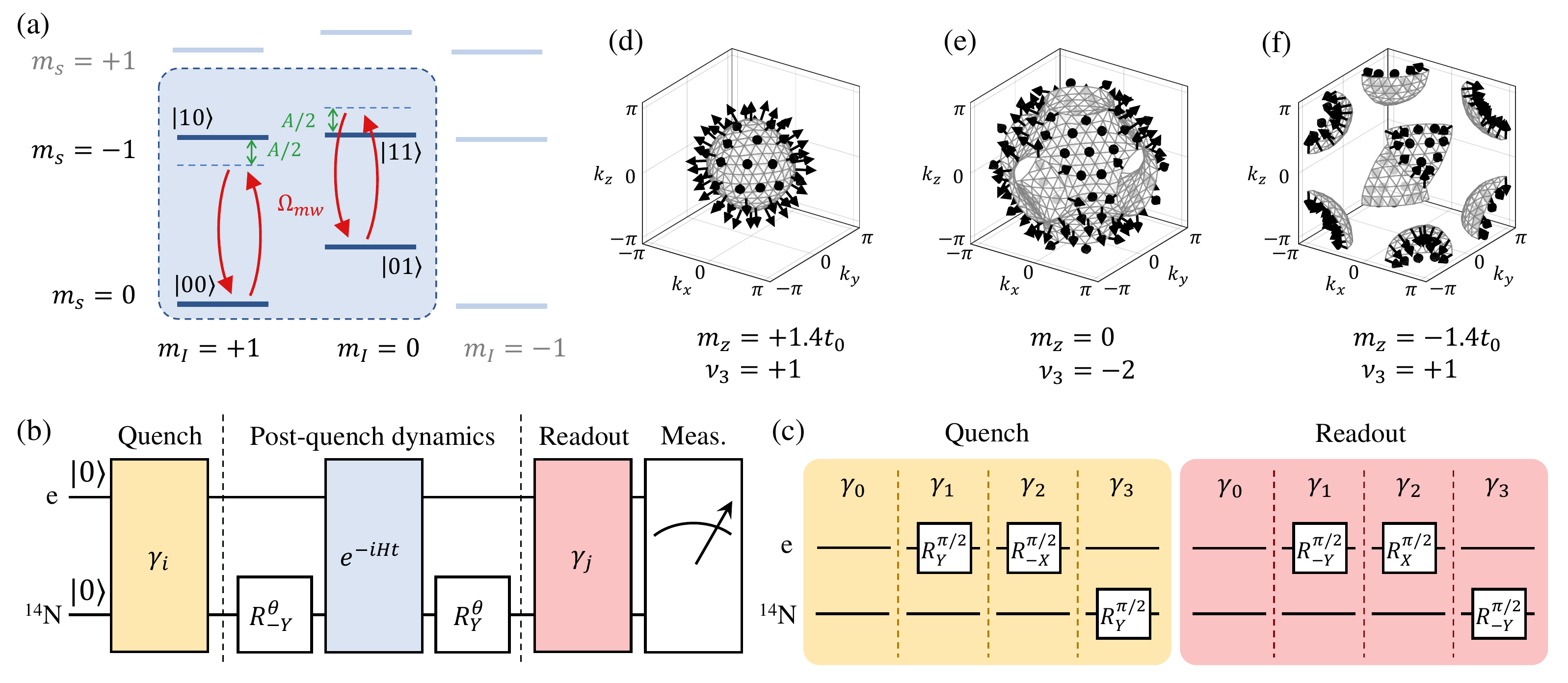}}
	\caption{
		(color online). \textbf{Experimental scheme and the dynamical bulk-surface correspondence.}
		(a) The energy diagram of the NV center used for simulation.  The electron and the nitrogen nuclear spins are used to form the two-qubit system. A microwave pulse is applied to simulate the 3D model.
		(b) Quantum circuit for the experiment. The operator $e^{-iHt}$ corresponds to evolution under $\mathcal H_{mw,{\rm RWA}}$, while the net effect together with the two nuclear spin operation $R ^\theta _{\pm Y}$ is equivalent to evolution under $\mathcal H_{\rm eff}$.
		(c) Operations of quench and readout steps in (b).
		(d-f) Experimental observation of the band inversion surfaces (BISs) and the dynamical spin-texture field $\textbf g (\textbf k)$ (arrows) for $m_z=+1.4$, $0$ and $-1.4$, respectively. The winding of ${\bf g}({\bf k})$ on BISs characterizes the 3D winding number $\nu_3$. Here $t_{\rm so}=0.2t_0$.
	}\label{fig:1}
\end{figure*}

Despite the considerable achievements, only a small portion of the broad classes of topological phases predicted in theory have been observed in condensed matter physics~\cite{Zhang2019,Vergniory2019,Tang2019}. Quantum simulation~\cite{Feynman1982}, as a state-of-art technique, provides a powerful way to explore exotic quantum phases beyond natural limits~\cite{Georgescu2014}. A number of exotic quantum systems, such as the two-dimensional (2D) Haldane model~\cite{Jotzu2014} and 2D spin-orbit (SO) coupled minimal model for quantum anomalous Hall phase~\cite{Liu2014b,Wu2016,Sun2018}, 1D chiral topological phase \cite{Atala2013,Liu2013,Song2018,Xie2019}, and 3D semimetals~\cite{Song2019,Tan2019} have been successfully realized in a controllable fashion with various quantum simulators including the ultracold atoms~\cite{Bloch2012}, photonic crystals~\cite{Aspuru-Guzik2012,Lu2014}, and solid-state qubit systems~\cite{Houck2012}. Nevertheless, in these studies, either the bulk or only the boundary physics of the simulated topological states can be well explored. For example, in ultracold atoms, it is convenient to measure the bulk topology but hard to simulate the boundary~\cite{Jotzu2014,Wu2016,Sun2018,Aidelsburger2015,note-1}. Therefore, a complete study of both the bulk and boundary topological physics, and their correspondence is challenging for quantum simulators.

In this work, we build up a quantum simulator using nitrogen-vacancy (NV) center to investigate 3D chiral topological insulator which was not accessible in solid systems, and demonstrate a complete simulation of the bulk and surface topological physics of the simulated chiral phase. This study is based on the recently proposed dynamical bulk-surface correspondence in momentum space~\cite{Zhanglin2018,Zhanglong2018-1,Zhanglong2019a,Zhanglong2019b,LZhou2018,Sun2018b,Wang2019,Yi2019}, which bridges the bulk topology of a $d$D equilibrium phase and the nontrivial quench dynamics emerging on $(d-1)$D momentum hypersurfaces called band inversion surfaces (BISs). The dynamical bulk-surface correspondence resembles the bulk-boundary correspondence in real space, and is easier to emulate than the latter, since the momentum space can be readily engineered for quantum simulators. This facilitates complete study of the simulated topological phases. In demonstrating the correspondence between the bulk and surface topological physics, we show in experiment the chiral symmetry protection of the 3D topological phase, and further measure the topological charges to directly characterize the bulk topology, with an emergent dynamical topological transition being observed.
%{\color{red}We study the bulk and surface topological physics and reveal their correspondence in three aspects:
%(i) We show that the chiral symmetry protection of the 3D bulk topology is clearly uncovered by emergent dynamical topology on BISs.
%(ii) We directly characterize the bulk topology by dynamically detecting the total charges enclosed by BISs.
%(iii) We observe an emergent topological transition when varying the quench depth, and show that this transition can be identified by
%either the deformation of the dynamical field on BISs, or the movement of topological charges.}

\emph{Simulation of the 3D model.}---The 3D chiral topological insulator simulated in the current experiment is described by the Bloch Hamiltonian ${\cal H}_{\rm 3D}({\bold k}) = \sum_{j=0}^4 h_j \gamma_j$ as
\begin{align}
		\mathcal H_{\rm 3D}(\bf k)= & \bigr[ m_z - t_0 ( \cos k_x + \cos k_y + \cos k_z )\bigr] \gamma_0\nonumber\\
		& + t_{\rm so} ( \sin k_x \gamma_1 + \sin k_y \gamma_2 + \sin k_z \gamma_3 ),
\end{align}
where the Bloch momentum ${\bold k} = (k_x, k_y, k_z)$, the Dirac matrices $\gamma_0 = \sigma_z \otimes \tau_z$, $\gamma_1 = \sigma_x \otimes \textbf 1$, $\gamma_2 = \sigma_y \otimes \textbf 1$, and $\gamma_3 = \sigma_z \otimes \tau_x$, with the Pauli matrices $\sigma_{x,y,z}$ and $\tau_{x,y,z}$ in the present simulator corresponding to the electron and nuclear spins, respectively.
%The Hamiltonian can be rewritten in the effective Zeeman form as $\mathcal H_{3D}=h_0 \gamma_0 + \sum_{i=1,2,3} h_i \gamma_i $.
The $ h_0(\bold k)$-term with the parameters $m_z$ and $t_0$ characterizes the dispersion of four uncoupled bands. %with $m_z$ denoting the effective magnetization and $t_0$ denoting the spin-conserved
The remaining part, written as $ {\bf h}_{\rm so}( {\bf k} ) = (h_1,h_2,h_3) $, represents a spin-orbit field which couples the four different bands, with $t_{\rm so}$ simulating the spin-flipped hopping coefficient. The Hamiltonian has a chiral symmetry defined by $\gamma_4=\sigma_z \otimes \tau_y$, hence it belongs to AIII class according to the Altland-Zirnbauer ten-fold symmetry classification~\cite{Chiu2016,AZ1997} and is characterized by 3D winding numbers in the equilibrium theory. The topological phases include three nontrivial regions: (i) $t_0<m_0<3t_0$ with winding number $\nu_3=1$; (ii) $-t_0<m_0<t_0$ with $\nu_3=-2$; and (iii) $-3t_0<m_0<-t_0$ with $\nu_3=1$. Beyond these regions the phase is trivial, and across the phase transition points the bulk gap is closed.

We realize the Hamiltonian $\mathcal H_{\rm 3D}$ by a quantum simulator built from NV center in diamond~\cite{Doherty2013}. The electrons around the center form an effective electron spin with a triplet ground state ($S=1$). Together with the intrinsic nitrogen-14 nuclear spin ($I=1$), it forms a coupled system, as depicted in Fig.~\ref{fig:1}(a). The Hamiltonian of the NV center is
\begin{equation}
	\mathcal H_{\rm NV} = 2\pi (D S_z^2 + \omega_e S_z + Q I_z^2 + \omega_n I_z + A S_z I_z),
\end{equation}
where $S_z$ ($I_z$) denotes the electron (nuclear) spin operator, $D=2.87\mathrm{GHz}$ is the electronic zero-field splitting, $Q=-4.95\mathrm{MHz}$ is the nuclear quadrupolar interaction, and $A=-2.16\mathrm{MHz}$ is the hyperfine interaction.
A magnetic field of $514\mathrm{G}$ is applied along the NV's symmetry axis, yielding an electron (nuclear) Zeeman splitting $\omega_e$ ($\omega_n$) of $1439\ \mathrm{MHz}$ ($154\ \mathrm{kHz}$).
The subspace of $\{m_s=0,-1\} \otimes \{m_i=+1,0\}$ is utilized to form a two-qubit system, %for the present quantum simulator, which is
relabeled as $\{\left|0\right> ,\left|1\right>\} \otimes \{\left|0\right> ,\left|1\right>\}$, on which the Pauli operators $\sigma_i$ and $\tau_i$ are defined. A microwave pulse is applied to produce an external driving field $\Omega_{mw}$.
Under the rotating-wave approximation, the effective Hamiltonian capturing the couplings in the subspace reads $\mathcal H_{mw,{\rm RWA}} =2\pi\left( \frac{A}{4}\sigma_z\tau_z +\Omega_x\sigma_x +\Omega_y\sigma_y \right)$, with $\Omega_x=\Omega_{mw} \cos \phi$ and $\Omega_y=-\Omega_{mw} \sin \phi$, where $\phi$ is the phase of microwave pulse. Finally, the $\sigma_z\tau_x$ term can be generated via a unitary rotation about $\tau_y$ axis by $\theta$ angle, yielding
\begin{eqnarray}
	\mathcal H_{\rm eff} = \frac{A}{4}\cos \theta \sigma_z\tau_z +\Omega_x\sigma_x
		+\Omega_y\sigma_y +\frac{A}{4}\sin \theta \sigma_z\tau_x,
\end{eqnarray}
where the factor $2\pi$ is neglected. This $\tau_y$-rotation of the Hamiltonian is realized by applying a radio-frequency pulse to rotate the nuclear spin. The experiment was performed on a home-built confocal setup at room temperature. We use a [111] oriented NV center with solid immersion lens. The MW and radio-frequency control of NV center are realized through an arbitrary wave generator.

The 3D chiral topological insulator model $\mathcal H_{\rm 3D}$ can emulated by $\mathcal H_{\rm eff}$ after mapping the parameter space to Bloch momentum space, i.e. $(\theta, \phi, \Omega_{mw})\rightarrow \bold k$. Any state evolving under $\mathcal H_{\rm eff}$ is then mapped to the one evolving under $\mathcal H_{\rm 3D}$~\cite{Supp}. Thus the $\bold k$-space of the 3D chiral phase can fully engineered, while the real space including the boundary cannot be simulated for the quantum simulator. The key observation is that, as studied below, the dynamical bulk-surface correspondence in momentum space provides the alternative full investigation of the bulk and surface topological physics.

\emph{Dynamical bulk-surface correspondence.}---We present the nontrivial quench dynamics emerging on BISs and connected to the bulk topology.
The experimental procedure for quench along $\gamma_i$ axis consists of three steps [see Fig.~\ref{fig:1}(b)]. First, we initialize the state to the state $\left | 00 \right >$, which is then prepared to be fully (or incompletely) polarized along $\gamma_i$ axis by a unitary control. Then, the initialized state evolves by time $t$ under ${\cal H}_{\rm 3D}$, as simulated by $\mathcal H_{\rm eff}$, rendering the quench dynamics. Finally, we measure the spin polarization $\langle\gamma_j(t)\rangle$.
The opposite unitary operations are respectively used to perform the quench and measurement with respect to all the spin components [Fig.~\ref{fig:1}(c)].
Following the measurement, we obtain the time-averaged spin polarizations $\overline{\left < \gamma_j \right >}_i=\lim_{T\rightarrow\infty}(1/T)\int_0^T\left < \gamma_j(t) \right >_idt\propto h_i h_j$, which are key ingredients to characterize the topology~\cite{Zhanglin2018}. Here the index $i$ ($j$) denotes the quench (measurement) axis.

The dynamical bulk-surface correspondence states that the bulk topology of the 3D chiral topological phase uniquely corresponds to the nontrivial quench dynamics emerging on the 2D BISs~\cite{Zhanglin2018}. For the initial state fully polarized in the axis $\gamma_0$, the 2D BISs are formed by all the momenta points where spin oscillations are {\em resonant} and easily measurable, giving the vanishing time-averaged spin-polarizations
\begin{align}
	\mathrm{BIS}=\{\mathbf{k}\vert\overline{\langle\gamma_{i}(\mathbf{k})\rangle}_{0}=0, \ i=0,1,2,3\}.
\end{align}
A dynamical invariant can be defined on the BISs as the winding of an emergent dynamical spin-texture field $\mathbf{g}(\mathbf{k})$, with the $i$-th component $\ g_i( \textbf k ) = \frac{1}{\mathcal N_k} \partial_{k_\perp} \overline{\left < \gamma_i \right >}_0\big\vert_{\scriptscriptstyle{\bf k}\in\rm BIS}$ describing the variation slope of $\overline{\left < \gamma_i \right >}_0$ along the local direction $k_\perp$ perpendicular to the BISs and normalized by ${\mathcal N_k}$, and
\begin{align}
	\mathcal{W} = \frac{1}{8\pi}\int_{\rm BIS}\mathrm{d}^{2}\mathbf{k}\,\mathbf{g}\cdot(\nabla\mathbf{g}\times\nabla\mathbf{g}),
\end{align}
Geometrically, the dynamical invariant describes the coverage of the dynamical field $\mathbf{g}(\mathbf{k})$ over a 2D spherical surface. This dynamical topological invariant equals the bulk topological invariant of the ground band of ${\cal H}_{\rm 3D}$, and provides the dynamical characterization of the 3D chiral phase~\cite{Zhanglin2018}.

We show the experimental measurements of the three different topological regimes in Fig.~\ref{fig:1}(d-f). The BISs are measured and exhibit very different shapes in different phases. The measured dynamical field $\mathbf{g}(\mathbf{k})$ is depicted as arrows, from which with the 2D dynamical invariant $\mathcal{W}$ can be computed, and is verified to characterize the 3D bulk topology.

\begin{figure}
	\centering
%	\fbox
	{\includegraphics[width=1.0\columnwidth]{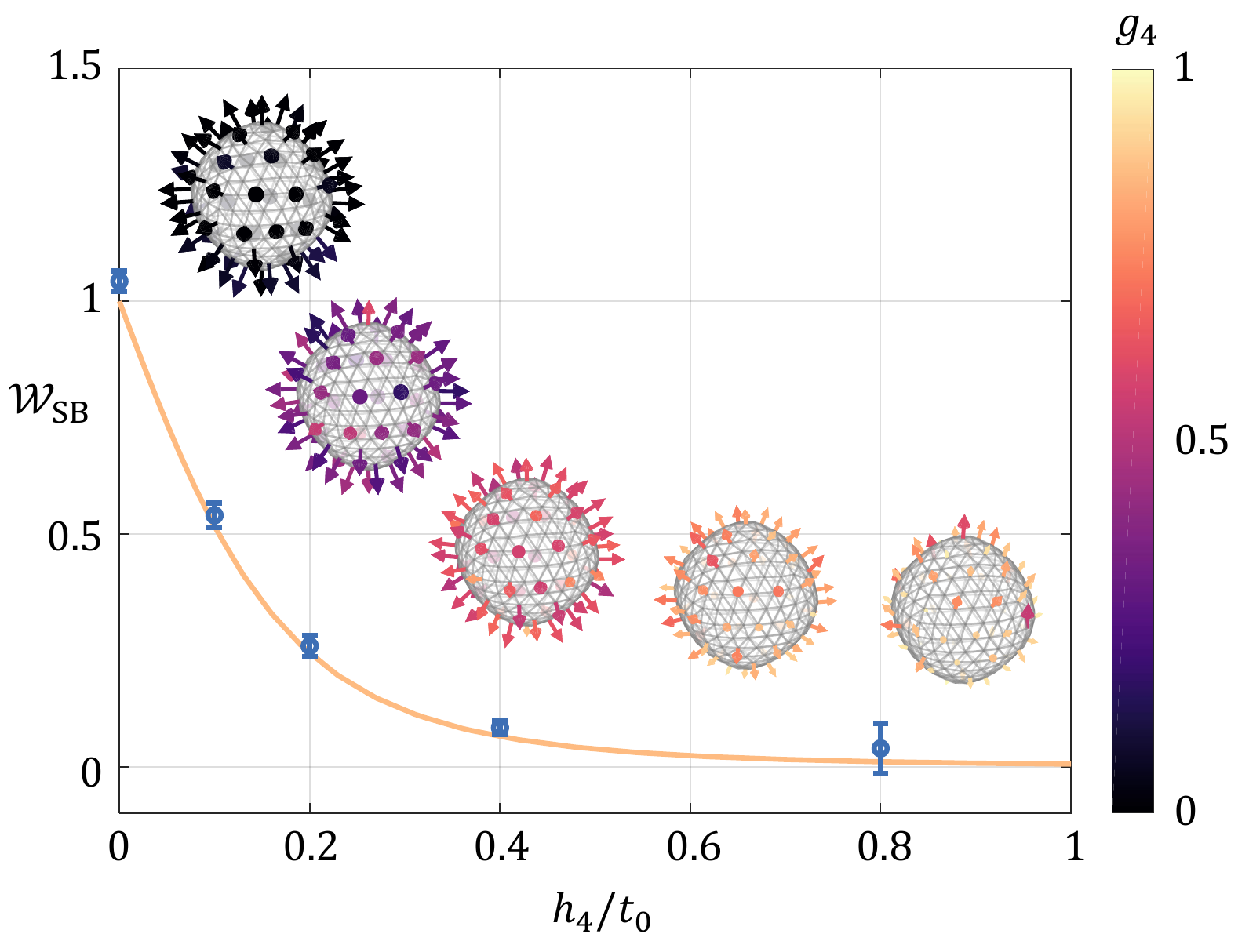}}
	\caption{
		(color online). \textbf{Measuring the chiral symmetry protection}. The blue data points are experimental results of winding numbers obtained from the emergent dynamical spin-texture field, and the error-bars represent three standard deviation. The orange line is calculated from the theoretic model. The insets are the emergent dynamical spin-texture field
$\mathbf{g}(\mathbf{k})=(g_1,g_2,g_3,g_4)$ correspond to each data point. The arrows denote $g_{1,2,3}$ components, and the color of the arrows denotes the component $g_4$.
	}\label{fig:sym_break}		
\end{figure}

\begin{figure*}
	\centering
%	\fbox
	{\includegraphics[width=1.8\columnwidth]{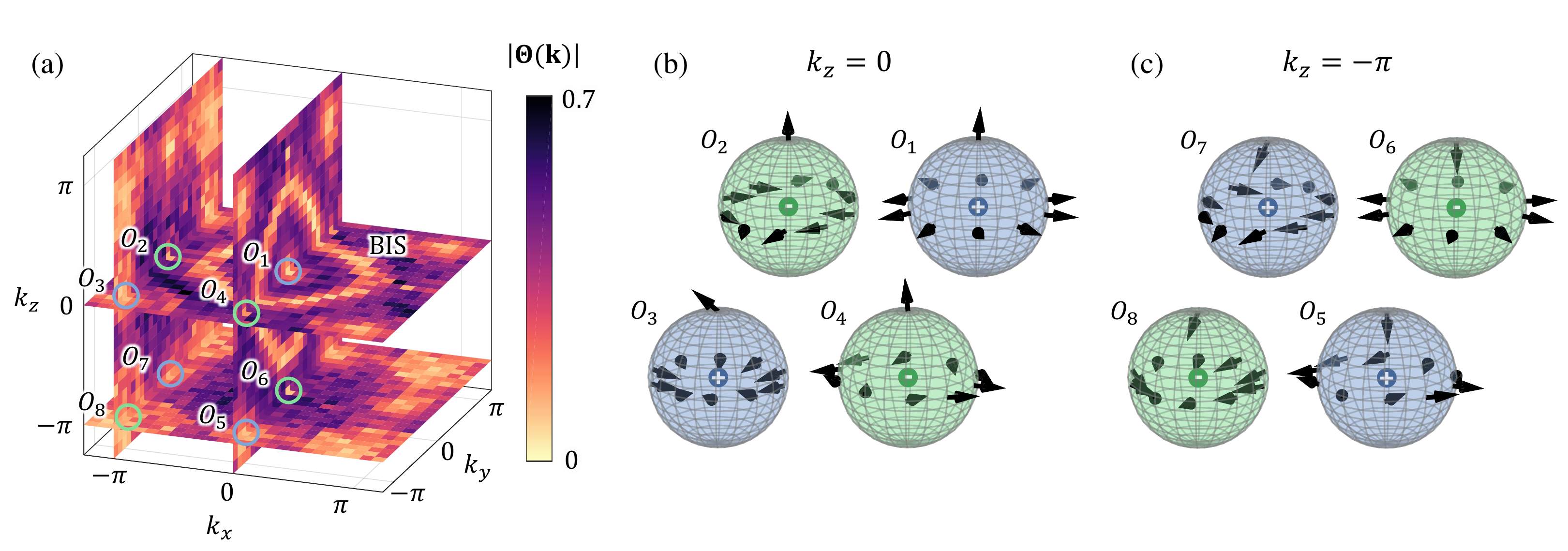}}
	\caption{
		(color online). \textbf{Dynamical measurement of topological charges.}
		(a) Experimental results of the dynamical field $\mathbf \Theta (\mathbf k)$ constructed from $\overline{\left < \gamma_0 (\textbf k) \right >}_{1,2,3}$. The color of the pixels indicates the norm of $\mathbf \Theta (\mathbf k)$, which vanishes at the location of BIS and topological charges. The positive (negative) charges are marked out with blue (green) circles.
		(b)(c) Dynamical field $\mathbf \Theta (\mathbf k)$ at around the charges on $k_z=0$ and $k_z=-\pi$ plane.
	}\label{fig:charge}
\end{figure*}

\emph{Measuring the chiral symmetry protection}.---The dynamical bulk-surface correspondence is protected by the chiral symmetry. It is important to verify the symmetry protection by studying the symmetry-breaking effect on the quench dynamics on BISs, similar to the symmetry-breaking in the boundary states in real space.
We create a constant term $h_{4}\gamma_4=h_4\sigma_z\tau_y$ into $\mathcal{H}_{\rm 3D}$ to break the chiral symmetry via an additional rotation in the $\tau_x$ axis. Then the dynamical spin-texture field on BISs becomes a 4D vector
$\mathbf{g}(\mathbf{k})=(g_1,g_2,g_3,g_4)$ with $g_i=\partial_{k_{\perp}}\overline{\langle\gamma_{i}\rangle}_{0}/\mathcal{N}_{\mathbf{k}}$,
%$\mathbf{g}(\mathbf{k})=(1/\mathcal{N}_{\mathbf{k}})(\partial_{k_{\perp}}\overline{\langle\gamma_{1}\rangle}_{0},\partial_{k_{\perp}}\overline{\langle\gamma_{2}\rangle}_{0},\partial_{k_{\perp}}\overline{\langle\gamma_{3}\rangle}_{0},\partial_{k_{\perp}}\overline{\langle\gamma_{4}\rangle}_{0})$,
which locates on a 3D spherical surface $S^3$. To quantify the geometric property of $\mathbf{g}(\mathbf{k})$, we note that without symmetry-breaking, i.e. $h_4=0$, the dynamical field $\mathbf{g}(\mathbf{k})$ sits on the equator of $S^3$ (equivalent to $S^2$). The solid angle enclosed by $\mathbf{g}(\mathbf{k})$ is a multiple of the half 3-sphere area, characterizing the invariant which can be generalized to the symmetry breaking case as
\begin{align}\label{W_CSB}
	\mathcal{W}_{\rm SB}=\frac{1}{\pi^2} \int_{\mathcal{S}\vert_{\partial\mathcal{S}=\textbf g(\textbf k)}} \textrm d S^3,
\end{align}
where $\pi^2$ is the area of the half unit 3-sphere, $\textrm d S^3$ is the corresponding area element,
and the integral is performed over the region $\mathcal{S}$ with boundary $\partial\mathcal{S}=\textbf g(\textbf k)$~\cite{Supp}.

The experimental measurement of the symmetry breaking effect is shown in Fig.~\ref{fig:sym_break}. It is observed that once $h_4\neq0$, the 4D dynamical field $\mathbf{g}(\mathbf{k})$ is shifted away from the equator of $S^3$, with a nonzero polarization in the $\gamma_{4}$ axis, for which the value $\mathcal{W}_{\rm SB}\neq\mathcal{W}$ is no longer quantized and decreases with the strength $h_4$.
The results show that the symmetry protection of the bulk topological phase can be identified from the the dynamical spin textures on BISs, which exhibit perfect (broken) topology and zero (nonzero) $\gamma_{4}$-polarization when the chiral symmetry is preserved (broken), similar to the boundary modes in real space which can be gapped out and polarized by the symmetry-breaking term.

\emph{Topological charges and emergent dynamical transition}.---We proceed to detect topological charges and characterize directly the bulk topology by the total charges enclosed by BISs~\cite{Zhanglong2019a}, which further demonstrates the correspondence between the bulk and surface topological physics. In this case, instead of measuring all spin components after a single quench along $\gamma_{0}$, we perform a series of quantum quenches along different $\gamma_{i}$ ($i=0,1,2,3$) axes but measure only $\gamma_{0}$ component, i.e.  $\overline{\langle\gamma_{0}(\mathbf{k})\rangle}_{i}$ after each quench~\cite{Zhanglong2019a,Zhanglong2019b}. To realize quenches in different axes $\gamma_{i}$, the quench process [see Fig.~\ref{fig:1}(c)] is modified to an appropriate combination of the nuclear and electron spin rotations, such that the resulting initial state is the eigenstate of the pre-quench Hamiltonian $\mathcal H_{\rm pre}=m_i \gamma_i + \mathcal H_{\rm 3D}$~\cite{Supp}. The BISs are again the collection of momenta on which time-averaged spin polarizations all vanish, namely $\mathrm{BIS}=\{\mathbf{k}\vert\overline{\left < \gamma_0( \textbf k ) \right >}_{i} =0,\forall i\}$, and the locations of topological charges are determined by $ \overline{\left < \gamma_0( \textbf k ) \right >}_{1,2,3} =0 $ while $\overline{\left < \gamma_0( \textbf k ) \right >}_0 \neq 0 $~\cite{Zhanglong2019a}. The topological charge is characterized by the dynamical field $\mathbf \Theta (\mathbf k)$ with components ($i=1,2,3$)
\begin{align}
	\Theta_i(\textbf{k})\equiv\frac{\mathrm{sgn}(h_0(\textbf{k}))}{\mathcal N_k} \overline{\left < \gamma_0( \textbf k ) \right >}_i.
\end{align}
The norm of $\mathbf \Theta (\mathbf k)$ vanishes at a charge, and the charge value equals the winding of the dynamical field $\mathbf \Theta (\mathbf k)$ near the charge. In Fig.~\ref{fig:charge}(a), we measure the norm of $|\mathbf \Theta (\mathbf k)|$ on the $k_{z}=0,-\pi$ and $k_{x}=0,-\pi$ planes for the phase with $m_z=1.4t_{0}$ and $t_{\rm so}=t_{0}$, which shows eight topological charges $O_{i}$ in the bulk. The dynamical field $\mathbf \Theta (\mathbf k)$ near the charges is shown in Fig.~\ref{fig:charge}(b) and (c), showing that the topological charges $O_{1,3,5,7}$ have value $+1$ while the charges $O_{2,4,6,8}$ have value $-1$. The bulk topology is characterized by the total topological charge enclosed by the BIS [see Fig.~\ref{fig:charge}(a)], which is $O_1$, giving $\nu_{3}=+1$ for the bulk phase [Fig.~\ref{fig:1}(d)].

\begin{figure}
	\centering
%	\fbox
	{\includegraphics[width=1\columnwidth]{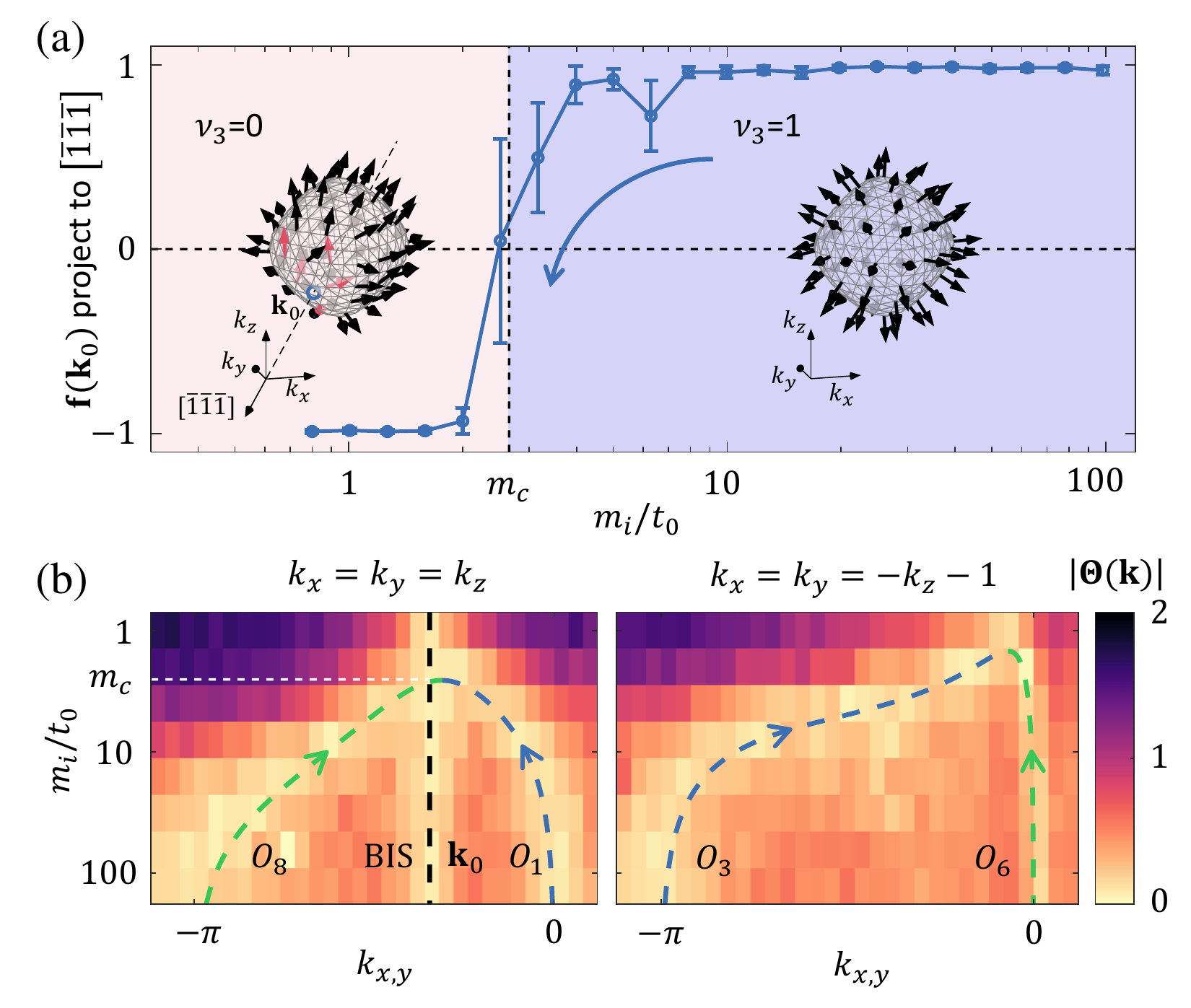}}
	\caption{
		(color online). \textbf{Identifying dynamical topological transition.}
		(a) Experimental results of dynamical spin-texture field $\textbf f(\textbf k_0)$ projected onto $[\overline 1 \overline 1 \overline 1]$ direction versus quench field. Here $k_{0x}=k_{0y}=k_{0z} = -1.084$ is the momentum on BIS in the $[\overline 1 \overline 1 \overline 1]$ direction. The emergent topological transition point is around $m_c = 2.7 t_0$. Insets are $\textbf f(\textbf k)$ for a shallow quench (left inset, $m_i = 2 t_0$) and a deep quench (right inset, $m_i = \infty$). At $\textbf k_0$ point (blue circle) in the left inset, $\textbf f(\textbf k)$ points inward the BIS (highlighted in red).
		(b) Locations of topological charges versus quench field. Dynamical field $\mathbf \Theta (\mathbf k)$ is measured along the line connecting charges $O_1$-$O_8$ (left) and $O_3$-$O_6$ (right), with the quench field $m_i$ being varied. %The BIS remains unchanged, while the charges move along the lines and merges at some point. Note that the charge $O_8$ merges with BIS at the emergent topological transition point $m_c$.
		The charge $O_8$ cross the BIS at $m_c$ and merges with $O_1$.
	}\label{fig:charge_moving}
\end{figure}

An interesting observation of the dynamical characterization with topological charges is that an emergent dynamical topological transition occurs when the quenches are varied from deep to shallow regimes~\cite{Zhanglong2019b}.
The deep (shallow) quench regime corresponds to large (small) $m_{i}$, and the initial state is fully (partially) polarized in the axis $\gamma_i$, independent of (dependent on) $\bold k$. For simplicity, in each set of quenches we take the same $m_{i}$ when quenching in different axes.
The BISs are not affected by the quench depth $m_i$~\cite{Zhanglong2019b}. To characterize the emergent topological transition we measure the dynamical field $\mathbf{f}(\mathbf{k})$ on BISs, with components given by
\begin{equation}
	f_i( \textbf k ) = \frac{1}{\mathcal N_k} \partial_{k_\perp} \overline{\left < \gamma_0 \right >}_i\big\vert_{\scriptscriptstyle{\bf k}\in\rm BIS},
\end{equation}
with $i=1,2,3$. For deep quenches, $\mathbf{f}(\mathbf{k})$ is equivalent to $\mathbf{g}(\textbf k)$, whose winding on BISs characterizes the post-quench topology [see the right insert in Fig.~\ref{fig:charge_moving}(a)]. In the left insert of Fig.~\ref{fig:charge_moving}(a), we show that the dynamical field $\mathbf{f}(\mathbf{k})$ for shallow quenches with $m_i=2$ is deformed and has zero winding, implying that between deep and shallow quenches an emergent topological transition occurs. To determine the critical quench depth $m_{c}$, we notice that the dynamical field changes most dramatically near the momentum $\mathbf{k}_{0}$ on the BIS in the $[\overline{111}]$ direction.
In reducing $m_i$ across $m_c$, the direction of $\mathbf{f}(\mathbf{k}_0)$ changes from the outward to inward of BIS, and vanishes at $m_i=m_c$, where the winding on BIS is ill-defined. Fig.~\ref{fig:charge_moving}(a) displays the projection of $\mathbf{f}(\mathbf{k}_0)$ in the $[\overline{111}]$ direction. Our measurement determines the critical value $m_{c}\backsimeq2.7t_{0}$, which agrees on the theoretical prediction $m_c=2.653t_{0}$.

The dynamical topological transition corresponds to the movement of topological charges across BIS, as illustrated in Fig.~\ref{fig:charge_moving}(b). The dynamical field $\mathbf{\Theta}(\mathbf{k})$ is measured along the line connecting charges $O_1$-$O_8$ or $O_3$-$O_6$. We observe that the locations of charges depend on the quench depth $m_{i}$. Particularly, the charge $O_8$ passes through BIS when reducing $m_i$ across the critical value $m_{c}$. Then no topological charge is enclosed by the BIS, providing the alternative picture of the emergent topological transition. The topological charges $O_3$ and $O_6$ can also annihilate at certain $m_{i}$, but do not change the dynamical topology on the BIS.

{\em Conclusion.---}In summary, we have achieved by quantum quenches a complete study of bulk and surface topological physics for a novel 3D chiral topological insulator, using a quantum simulator built from solid-state spin system. We experimentally identified the dynamical bulk-surface correspondence in momentum space, as a momentum-space counterpart of the bulk-boundary correspondence in real space, which bridges the bulk topology of the 3D chiral phase and the nontrivial quench dynamics emerging on 2D band inversion surfaces. As the momentum space is more convenient to engineer for quantum simulators, the dynamical bulk-surface correspondence enables a complete study of the simulated topological phases, without the necessity of constructing real-space boundaries. The novel topological physics have been observed in experiment, including the chiral symmetry protection, the topological charges, and the dynamical topological transition emerging in the quench studies. The present work showed the insightful techniques of quantum simulation, which can be easily extended to other simulators, and opens a broad avenue to explore high dimensional topological phases beyond the limits of condensed matter physics.

\emph{Acknowledgement.-}
This work is supported by
the National Key R$\&$D Program of China (Grant No. 2018YFA0306600, 2017YFA0305000, 2016YFA0301604, 2016YFB0501603),
the NNSFC (Grants No. 11775209, 11825401, 81788101, 11761161003, 11921005, 11761131011, 11722544),
the CAS (Grants No. GJJSTD20170001, No. QYZDY-SSW-SLH004, No. QYZDB-SSW-SLH005),
Anhui Initiative in Quantum Information Technologies (Grant No. AHY050000),
the Fundamental Research Funds for the Central Universities, the Thousand-Young-Talent Program of China.
%\textbf{Author contributions:} J.D.supervised the project; X.-J.L.,Y.W. and J.D. conceived the ideas; L.Z. and X.-J.L. formulated the theory; Y.W. supervised the experiments;
%W.T.J., M.Q.W. Y.G. and Z.H.C. prepared the sample and performed the experiments; W.T.J. and L.Z. wrote the manuscript with input from all authors. All authors along reviewed the manuscript and suggested improvements.
%\textbf{Competing Interests:} All authors declare that they have no competing interests.
%\textbf{Data and materials availability:} All data needed to evaluate the conclusions in the paper are present in the paper and/or the Supplementary Materials. Additional data related to this paper may be requested from the authors.
% \bibliography{references}

\begin{thebibliography}{99}
\bibitem{note-1} A synthetic boundary can be simulated with the synthetic dimension formed by internal degree of freedom, e.g. the hyperfine levels of ultracold atoms~\cite{synthetic1,synthetic2}.
\end{thebibliography}

%%%%%%%%%%%%%%%%%%%%%%%%%%%%%%%%%%%%%%%%%%%%%

\clearpage

\section{\textbf{\large{Supplementary material}}}

%\tableofcontents
%
\renewcommand{\thefigure}{A\arabic{figure}}

\setcounter{figure}{0}

\section{Generalized topological invariant}

As described in the main text, in the presence of the symmetry-breaking term $h_4 \gamma_4$, the generalized topological invariant is defined on a 3-sphere as
\begin{equation}
\mathcal{W}_{\rm SB}=\frac{1}{\pi^2} \int_{\mathcal{S}\vert_{\partial\mathcal{S}=\textbf g(\textbf k)}} \mathrm{d} S^3,
\end{equation}
where $\pi^2$ is the area of the half 3-sphere, $\mathrm{d} S^3$ is the area element of the 3-sphere, and the integral is taken over the region bounded by the curve $\partial\mathcal{S}=\textbf g(\textbf k)$.
%where the space $\mathcal{S}$ is four dimensional spanned by $\textbf g(\textbf k)=(g_1, g_2, g_3, g_4)$, and the integral is taken over the region bounded by the curve $\partial\mathcal{S}=\textbf g(\textbf k)$.
When there is no symmetry-breaking term $\gamma_4$, the dynamical field $\mathbf g (\mathbf k)$ lies on the equator (a 2D object on $S^3$), and the integral $\int_{\mathcal{S}\vert_{\partial\mathcal{S}=\textbf g(\textbf k)}} \mathrm{d} S^3$ is taken over the upper-half sphere and equals $n \pi^2$ if the dynamical field $\mathbf g (\mathbf k)$ winds the equator $n$ times, giving the bulk topological invariant $\mathcal W = n$. When the symmetry is broken, the dynamical field $\mathbf g (\mathbf k)$ deviates from the equator, the integral $\int_{\mathcal{S}\vert_{\partial\mathcal{S}=\textbf g(\textbf k)}} \mathrm{d} S^3,$ generally is a fraction multiplying $\pi^2$, then $\mathcal W$ is a non-integer number as shown below.

In the sphere coordinates, we have
\begin{equation}
\mathrm{d} S^3 = \sin^2 \phi_3 \sin \phi_2 \mathrm{d} \phi_1 \mathrm{d} \phi_2 \mathrm{d} \phi_3,
\end{equation}
where
\begin{equation}
\begin{aligned}
g_4 & = \cos \phi_3,\\
g_3 & = \sin \phi_3 \cos \phi_2,\\
g_2 & = \sin \phi_3 \sin \phi_2 \cos \phi_1,\\
g_1 & = \sin \phi_3 \sin \phi_2 \sin \phi_1
\end{aligned}
\end{equation}
with $0<\phi_1<2\pi$ and $0<\phi_{2,3}<\pi$. Inversely, we have
\begin{equation}
\begin{aligned}
\tan \phi_1 & = g_1/g_2,\\
\tan \phi_2 & = \sqrt{g_1^2+g_2^2}/g_3,\\
\tan \phi_3 & = \sqrt{g_1^2+g_2^2+g_3^2}/g_4.
\end{aligned}
\end{equation}
If the symmetry is not broken, then $g_4=0$ and $\phi_3=\pi/2$, the dynamical field lies on the equator of the 3-sphere. Here we consider the case where the dynamical field $\mathbf g (\mathbf k)$ winds the equator $n$ times, i.e., the bulk topological invariant is $n$, then we have
\begin{equation}
\begin{aligned}
\mathcal W_{\rm SB} & = \frac{1}{\pi^2} \int_{ \mathcal{S} \vert _{\partial \mathcal{S} = \textbf g(\textbf k)}} \mathrm{d} S^3 \\
& =  \frac{1}{\pi^2} \int_{0}^{2\pi n} \mathrm{d} \phi_1 \int_{0}^{\pi} \mathrm{d} \phi_2 \int_{0}^{\pi/2} \mathrm{d} \phi_3 \sin^2 \phi_3 \sin \phi_2 \\
& = n,
\end{aligned}
\end{equation}
Now we add the symmetry-breaking term, and we consider the following simple example, $g_4(\textbf k)=m$, where $|m|<1$ is a constant. Then the quantity $\mathcal W_{\rm SB}$ becomes
\begin{equation}\label{eq:winding}
\begin{aligned}
\mathcal W_{\rm SB} & =  \frac{1}{\pi^2} \int_{0}^{2\pi n} \mathrm{d} \phi_1 \int_{0}^{\pi} \mathrm{d} \phi_2 \int_{0}^{\arccos m} \mathrm{d} \phi_3 \sin^2 \phi_3 \sin \phi_2 \\
& = \frac{2n}{\pi}(\arccos m - m\sqrt{1-m^2}),
\end{aligned}
\end{equation}
which is not an integer.

\begin{figure}[t]
	\centering
	%	\fbox
	{\includegraphics[width=1.0\columnwidth]
		{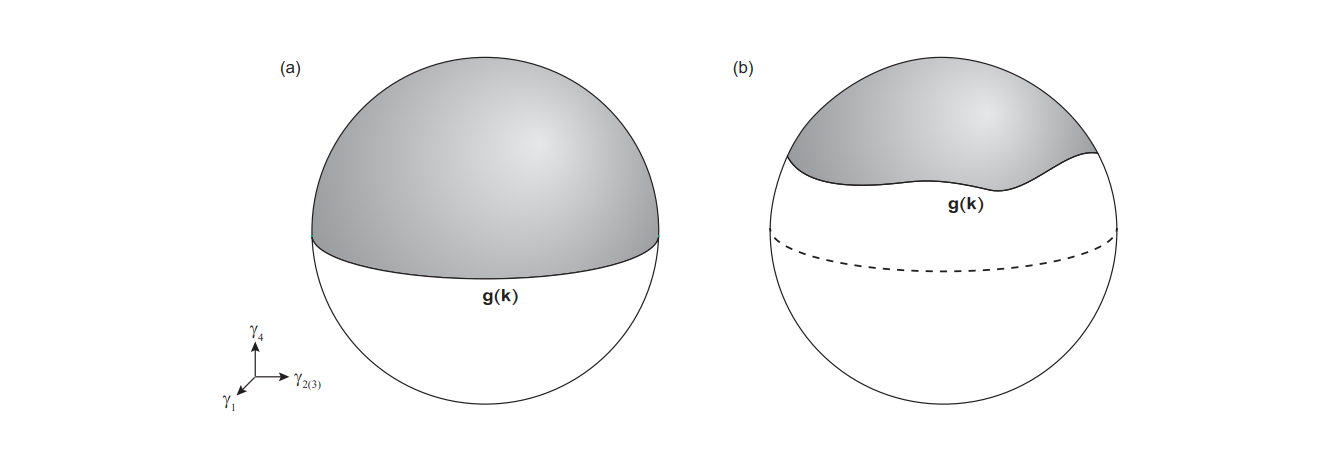}}
	\caption{
		\textbf{Illustration for the quantity $\mathcal W$ proportional to the area of the shadow region bounded by the dynamical field $\mathbf g (\mathbf k)$.}
		(a) is the case with the chiral symmetry, the dynamical field $\mathbf g (\mathbf k)$ lies on the equator.
		(b) is the case with the symmetry broken by the $\gamma_4$ term, and the dynamical field deviates from the equator. Note that the equator is now a 2D manifold on the 3-sphere.
	}
	\label{fig:winding}
\end{figure}

\section{Experimental simulation of the 3D model}

\subsection{Post-quench dynamics}
\label{sec:sys.ham}

The target Hamiltonian $\mathcal H_{\rm 3D}$ can be written in a general form
\begin{equation}
\mathcal H_{\rm 3D}(\textbf k) = h_0 \sigma_z \tau_z + h_1 \sigma_x + h_2 \sigma_y + h_3 \sigma_z \tau_x.
\end{equation}
We realize this Hamiltonian with a diamond nitrogen-vacancy (NV) center system, whose Hamiltonian is
\begin{equation}
\mathcal H_{NV} = 2\pi (D S_z^2 + \omega_e S_z + Q I_z^2 + \omega_n I_z + A S_z I_z),
\end{equation}
where $S_z$ ($I_z$) is the electron (nuclear) spin operator. A subspace of $\{m_s=0,-1\} \otimes \{m_i=+1,0\}$ is utilized to form a two-qubits system, which is relabeled as $\{\left|0\right> ,\left|1\right>\} \otimes \{\left|0\right> ,\left|1\right>\}$. The first (second) qubit corresponds to the Pauli operator $\sigma_i$ ($\tau_i$) in $\mathcal H_{\rm 3D}$. The subspace Hamiltonian can be rewritten as
\begin{equation}
H_0 = 2\pi
\left( \begin{array}{cccc}
\omega_1 	& 0        & 0        & 0        \\
0       & \omega_2 & 0        & 0        \\
0       & 0        & \omega_3 & 0        \\
0       & 0        & 0        & \omega_4
\end{array} \right),
\end{equation}
where $\omega_1=Q+\omega_n$, $\omega_2=0$, $\omega_3=D-\omega_e+Q+\omega_n-A$ and $\omega_4=D-\omega_e$.

In order to introduce $\sigma_x$ and $\sigma_y$ terms in $\mathcal H_{\rm 3D}$, we apply a microwave pulse of frequency $\omega_{mw}=(\omega_3-\omega_1+\omega_4-\omega_2)/2=D-\omega_e-A/2$, coupling both $\left|00\right> \leftrightarrow \left|10\right>$ and $\left|01\right> \leftrightarrow \left|11\right>$ transitions. The interaction Hamiltonian reads
\begin{equation}
V_{mw} = 2\pi \Omega_{mw} \cos(\omega_{mw} t+\phi)
\left( \begin{array} {cccc}
0 & 0 & 1 & 0 \\
0 & 0 & 0 & 1 \\
0 & 0 & 0 & 0 \\
0 & 0 & 0 & 0
\end{array} \right)+h.c.
\end{equation}
After transforming the total Hamiltonian $H_0+V_{mw}$ to the rotating frame defined by the MW field, and applying proper rotating-wave approximation, the system Hamiltonian reads
\begin{equation}
\begin{aligned}
\mathcal H_{mw,\rm RWA} &= 2\pi
\left( \begin{array} {cccc}
A/4 & 0 & \Omega_x-\mathrm i \Omega_y & 0 \\
0 & -A/4 & 0 & \Omega_x-\mathrm i \Omega_y \\
\Omega_x+\mathrm i \Omega_y & 0 & -A/4 & 0 \\
0 & \Omega_x+\mathrm i \Omega_y & 0 & A/4
\end{array} \right)\\
&=2\pi\left( \frac{A}{4}\sigma_z\tau_z +\Omega_x\sigma_x +\Omega_y\sigma_y \right),
\end{aligned}
\end{equation}
where $\Omega_x=\Omega_{mw} \cos \phi$ and $\Omega_y=-\Omega_{mw} \sin \phi$. $\mathcal H_{mw,\rm RWA}$ is only a single $\tau_y$ rotation away from the target Hamiltonian form $\mathcal H_{\rm 3D}=h_0 \sigma_z \tau_z + h_1 \sigma_x + h_2 \sigma_y + h_3 \sigma_z \tau_x$. After applying the rotation $U_{\rm rot}=\exp(-\mathrm i \theta \tau_y)$ to the system Hamiltonian, we have our effective Hamiltonian
\begin{equation}
\mathcal H_{\rm eff}=2\pi\left(\frac{A}{4}\cos \theta\ \sigma_z\tau_z +\Omega_x\sigma_x +\Omega_y\sigma_y +\frac{A}{4}\sin \theta\ \sigma_z\tau_x\right).
\end{equation}
We can imply that
\begin{equation}\label{eq:map1}
\theta=\arctan(h_3/h_0).
\end{equation}
Note that $\mathcal H_{\rm eff}$ is subject to the limitation of having $h_0^2+h_3^2=\pi^2 A^2 /4$. In order to simulate $\mathcal H_{\rm 3D}$ with any $h_i$, we here define the effective time as a rescale of the simulation time $t$, i.e., $ t_{\rm eff} = \alpha  t$. We only need to reproduce the same effect as $U_{3D}=\exp( -\mathrm i \mathcal H_{\rm 3D}  t  )$ with the simulated evolution $U_{\rm eff}=\exp( -\mathrm i \mathcal H_{\rm eff}  t_{\rm eff} )$. Hence we have $\mathcal H_{\rm 3D} = \alpha \mathcal H_{\rm eff} $, giving
\begin{equation}\label{eq:map2}
\begin{aligned}
\alpha      & =\frac{2}{\pi|A|} \sqrt{h_0^2+h_3^2},   \\
\Omega_{mw} & =\frac{1}{2\pi \alpha} \sqrt{h_1^2+h_2^2}, \\
\phi   & =-\arctan(h_2/h_1).
\end{aligned}
\end{equation}

\begin{figure}[t]
	\centering
	%	\fbox
	{\includegraphics[clip, trim=0cm 12cm 14cm 0cm, width=0.8\columnwidth]
		{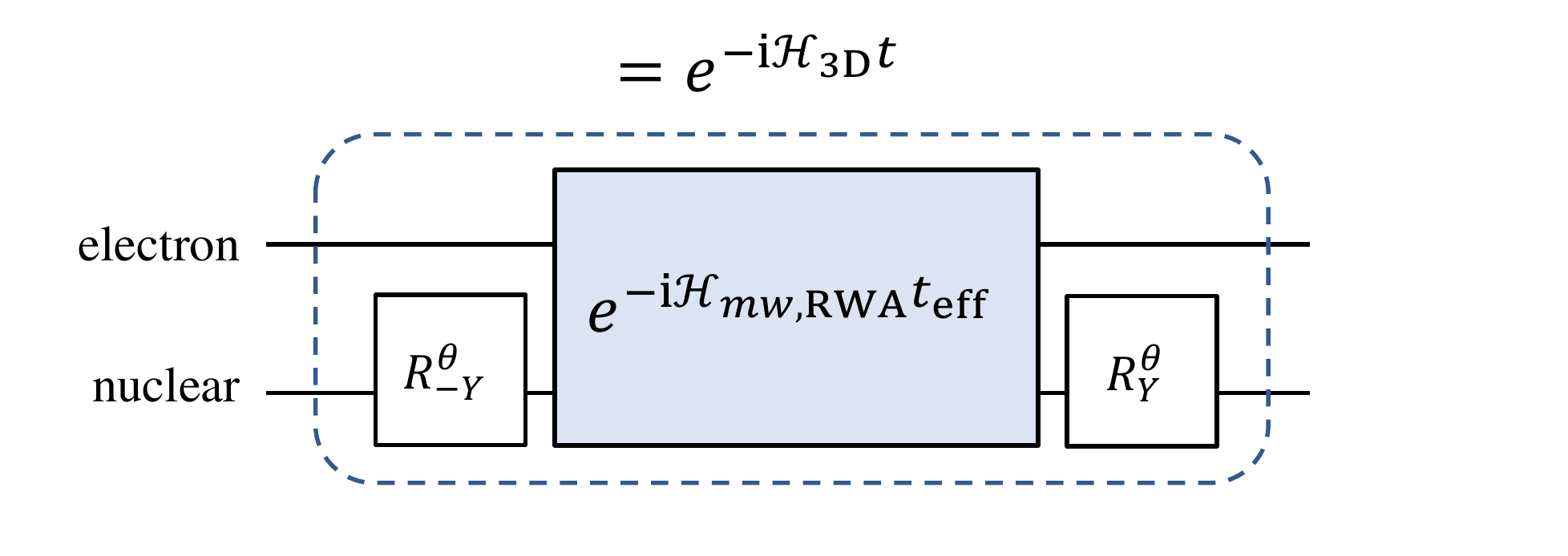}}
	\caption{
		\textbf{Post-quench dynamics.}
		$R^\theta_{\pm Y}$ represents rotation of nuclear spin about axis $\pm y$ for an angle $\theta$.
		The operation in the middle is evolution of the system under the microwave driving $\Omega_{mw}$, for a time duration of $ t_{\rm eff}$.
		The net effect is equivalent to evolution under $\mathcal H_{\rm 3D}$ for a duration of $ t $.
	}
	\label{fig:evol}
\end{figure}

The corresponding experimental circuit of this post-quench evolution is depicted in Fig.~\ref{fig:evol}. We first rotate the nuclear spin along $-y$ axis for an angle $\theta$. Then the microwave with a driving strength of $\Omega_{mw}$ and a phase of $\phi$, for a time duration of $ t_{\rm eff}$ is applied. Finally we rotate back the nuclear spin along $y$ axis for the same angle $\theta$. The net effect of this whole process is identical to the evolution of the system under $\mathcal H_{\rm 3D}$ during an evolution time of $ t $.

For the symmetry breaking case with the additional $\gamma_4$ term, the rotation operation $U_{\rm rot}$ is modified to $U_{\rm rot,SB}=\exp [-\mathrm i \theta (\cos \phi_{\rm SB} \tau_y - \sin \phi_{\rm SB} \tau_x)] $, which will give the post-rotation Hamiltonian as
\begin{equation}
\begin{aligned}
\mathcal H_{\rm sim,SB}= & 2\pi\left( \frac{A}{4}\cos \theta\ \sigma_z\tau_z
+\Omega_x\sigma_x -\Omega_y\sigma_y \right. \\
&\hspace{0.35cm} \left. +\frac{A}{4}\sin \theta \cos \phi_{\rm SB}\ \sigma_z\tau_x
+\frac{A}{4}\sin \theta \sin \phi_{\rm SB}\ \sigma_z\tau_y \right).
\end{aligned}
\end{equation}
This modified rotation is realized by setting the phase of the RF pulse to \begin{equation}
\phi_{\rm SB}=\arctan(h_4/h_3).
\end{equation}
$\theta$ and $\phi$ are also modified by substituting $h_3$ with $\sqrt{h_3^2+h_4^2}$, giving
\begin{equation}
\begin{aligned}
\theta & =\arctan(\sqrt{h_3^2+h_4^2}/h_0),\\
\alpha & =\frac{2}{\pi|A|} \sqrt{h_0^2+h_3^2+h_4^2}.
\end{aligned}
\end{equation}
Note that when setting $h_4=0$, we have $\phi_{\rm SB}=\pi$, $U_{\rm rot,SB}$, $\theta$ and $\alpha$ reduce to the non-symmetry-breaking case.

\subsection{Deep and shallow quench process}
\label{sec:quench}

In the experiment the system is initially polarized by a green laser pulse to the state $\left| 00 \right>$, which is an eigenstate of $\gamma_0 =\sigma_z \tau_z$. This is equivalent to a deep quench along $\gamma_0$. The deep quenching along other axis is realized by either applying a microwave or radio-frequency pulse to prepare the system onto the eigenstate of $\gamma_{1,2,3}$.

For the shallow quench process, one need to initialize the state to the eigenstate of the quench Hamiltonian $\mathcal H_{\rm pre}=m_i \gamma_j + \mathcal H_{\rm 3D}$, with a finite quench field $m_i$ along quench axis $\gamma_j$. In general, the quench Hamiltonian can be rewritten as
\begin{equation}
\mathcal H_{\rm pre} = h_{\rm pre,0} \sigma_z \tau_z + h_{\rm pre,1} \sigma_x + h_{\rm pre,2} \sigma_y + h_{\rm pre,3} \sigma_z \tau_x.
\end{equation}
which denpends on both $m_i$ and $\textbf k$, and is no longer aligned with any of the $\gamma_i$ axes. To prepare an eigenstate of $\mathcal H_{\rm pre}$, we consider a rotation of electron spin
\begin{equation}
\begin{aligned}[t]
	U_{\rm init,mw} &= \exp \left( -\mathrm i \frac{\theta_{\rm init,mw}}{2} (-\sin \phi_{\rm init,mw} \sigma_x +\cos \phi_{\rm init,mw} \sigma_y) \right),\\
	\theta_{\rm init,mw}  &= \arctan \left( \sqrt{h_{\rm pre,1}^2+h_{\rm pre,2}^2}/\sqrt{h_{\rm pre,0}^2+h_{\rm pre,3}^2} \right),\\
	\phi_{\rm init,mw}  &= \arctan \left( h_{\rm pre,2}/h_{\rm pre,1} \right).
\end{aligned}
\end{equation}
After the rotation, the state $U_{\rm init,mw} \left| 00 \right>$ becomes an eigenstate of $\sqrt{ h_{\rm pre,0}^2 + h_{\rm pre,3}^2 } \sigma_z + h_{\rm pre,1} \sigma_x + h_{\rm pre,2} \sigma_y$. Due to the fact that $U_{\rm init,mw}$ operates only on electron spin, it commutes with $\tau_z$. As a result, $U_{\rm init,mw} \left| 00 \right>$ is also an eigenstate of $\sqrt{ h_{\rm pre,0}^2 + h_{\rm pre,3}^2 } \sigma_z \tau_z + h_{\rm pre,1} \sigma_x + h_{\rm pre,2} \sigma_y$.
We can further rotate the nuclear spin as
\begin{equation}
\begin{aligned}[t]
	U_{\rm init,rf} &= \exp \left( -\mathrm i \frac{\theta_{\rm init,rf}}{2} \tau_y \right),\\
	\theta_{\rm init,rf}  &= \arctan \left( h_{\rm pre,3} / h_{\rm pre,0} \right).\\
\end{aligned}
\end{equation}
With this rotation, we end up with an eigenstate of $\mathcal H_{\rm pre}$. Note that $U_{\rm init,rf}$ commutes with $U_{\rm init,mw}$, the order of these two operations can be switched in the experiment.

\subsection{Readout time-averaged spin polarization}

\begin{figure}[t]
	\centering
%	\fbox
	{\includegraphics[clip, trim=2cm 7cm 2cm 0cm, width=1.0\columnwidth]
		{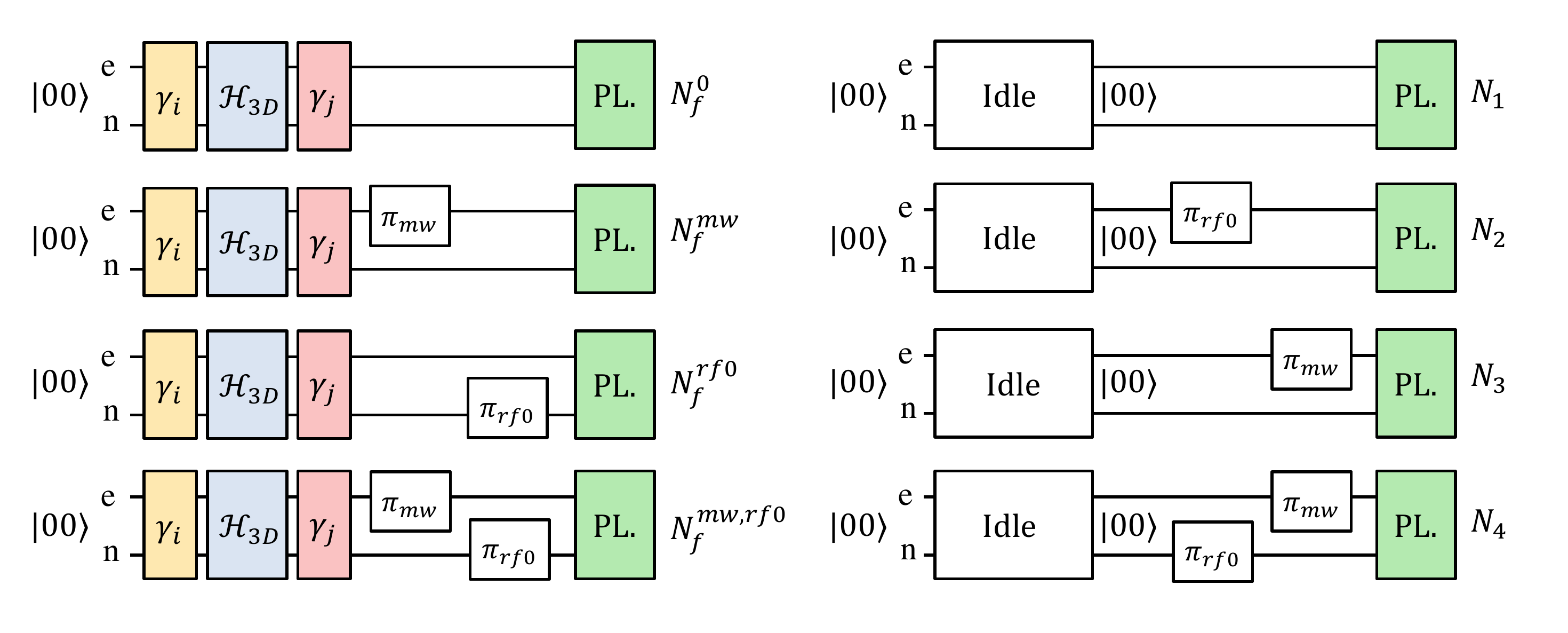}}
	\caption{
		\textbf{Sequences to measure populations.}
		The operation labeled $\gamma_i$ (yellow) corresponds to quench along $\gamma_i$ axis, which requires an operation to transform the state $\left| 00 \right>$ to an eigenstate of the quench Hamiltonian $\mathcal H_{\rm pre}$.
		The operation $\mathcal H_{\rm 3D}$ (blue) corresponds to the post-quench dynamics as depicted in Fig.~\ref{fig:evol}.
		The operation $\gamma_j$ (red) corresponds to readout $\gamma_j$ component, which requires an operation to transform $\gamma_j$ to the $z$ basis.
		PL. (green) corresponds to a photoluminescence measurement, which is realized by applying a 532nm laser and counting the emitted photons.
		Idle corresponds to a waiting time equal to the total time of the $\gamma_i$, $\mathcal H_{\rm 3D}$ and $\gamma_j$ steps.
	}
	\label{fig:tomo}
\end{figure}

\begin{figure}[t]
	\centering
%	\fbox
	{\includegraphics[clip, trim=2cm 7.5cm 2cm 0cm, width=1.0\columnwidth]
		{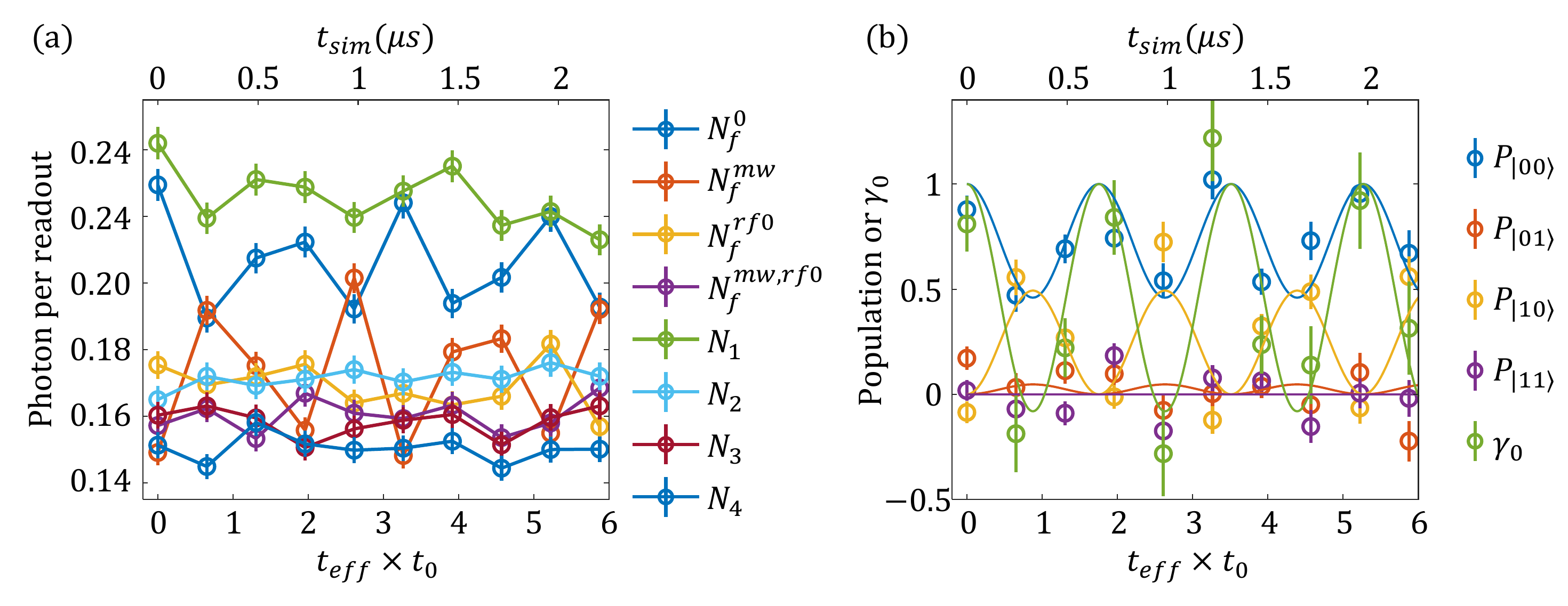}}
	\caption{
		\textbf{A typical measurement result.}
		(a) Photon number per readout for different readout sequences as shown in Fig.~\ref{fig:tomo} at different time. Error bars are estimated by photon shot noise.
		(b) Population and spin polarization calculated from results in (a). The time-averaged spin polarization calculated from this result is $ \overline{\left < \gamma_0 \right >}_0 = 0.423 \pm 0.056 $, comparing to the theory value of 0.460.
		For both (a) and (b), we set $m_z=1.4 t_0$, $t_{so}=0.2 t_0$, and the momentum point is $(k_x,k_y,k_z)=(0.1\pi,0.6\pi,0.1\pi)$, with deep quench along $\gamma_0$.
	}
	\label{fig:tomo_demo}
\end{figure}

The spin polarizations $\gamma_i$ of a given final state is measured by transforming the interested component to the $z$ basis of electron and nuclear spins, followed by a population measurement($P_{\left| i,j \right>}(i,j=0,1)$) through the optical readout.
For the $\gamma_0=\sigma_z \tau_z$ readout, the spin polarization is essentially $P_{\left| 00 \right>} -P_{\left| 01 \right>} -P_{\left| 10 \right>} +P_{\left| 11 \right>}$, which is already in $z$ basis.
For the case of $\gamma_{1,2} = \sigma_{x,y} \otimes \textbf 1$, a $\pi/2$ rotation on the electron spin about $-y$ or $x$ axis will map the $\sigma_{x,y}$ components to $\sigma_z$, of which the spin polarization is given by $P_{\left| 00 \right>} +P_{\left| 01 \right>} -P_{\left| 10 \right>} -P_{\left| 11 \right>}$.
Similarly, for the $\gamma_{3,4} = \sigma_z \tau_{x,y}$ readout, a $\pi/2$ rotation on the nuclear spin about $-y$ or $x$ axis will transform the $\gamma_{3,4}$ readout to a $\gamma_0$ readout. These operations are depicted in Fig.~1(c) of the main text.

For the populations readout, the photoluminescence (PL) photon count of the spin state is recorded. Since the total PL count is the average of all four levels weighted by their populations, i.e., $N_{total}= N_1 P_{\left| 00 \right>} +N_2 P_{\left| 01 \right>} +N_3 P_{\left| 10 \right>} +N_4 P_{\left| 11 \right>}$, we apply RF and MW pulses in different ways to produce different linear combinations of the populations, and then combine all the equations to solve for the populations. The sequences are depicted in Fig.~\ref{fig:tomo}, and the system of equations for the populations is
\begin{equation}
	\left( \begin{array} {cccc}
			N_1 & N_2 & N_3 & N_4 \\
			N_3 & N_4 & N_1 & N_2 \\
			N_2 & N_1 & N_3 & N_4 \\
			N_3 & N_4 & N_2 & N_1
		\end{array} \right)
		\cdot \left( \begin{array} {c}
			P_{\left| 00 \right>} \\
			P_{\left| 01 \right>} \\
			P_{\left| 10 \right>} \\
			P_{\left| 11 \right>}
		\end{array} \right)
		= \left( \begin{array} {l}
			N_f^0 \\
			N_f^{mw} \\
			N_f^{rf0} \\
			N_f^{mw,rf0}
	\end{array} \right).
\end{equation}
Note that the $N_{1,2,3,4}$ also need to be determined. The sequences are also depicted in Fig.~\ref{fig:tomo}.

The time-averaged spin polarization $\overline{ \left< \gamma_i( \textbf{k} ) \right> }$ is obtained by measuring and averaging spin polarization over a series of time. In order to maintain consistency, the time steps are chosen in such a way that the corresponding simulation time $ t $ are the same in all comparable measurements. Note that since the effective time $ t_{\rm eff}$ also depends on $\textbf{k}$, the same $ t $ may correspond to different $ t_{\rm eff}$. For experiments in which the effect of dephasing is ignored, the time range of $ t $ is chosen from 0 to
\begin{equation}
	t_{\rm max}=\frac{2}{\sqrt 3 \ t_{\rm so} \sin{ (\arccos{ (3 m_z/t_0) } )}}.\\
\end{equation}\label{eq:t_eff_max}
For the experiments with dephasing, the time range is chosen from $2 t_{\rm max}$ to $3 t_{\rm max}$.
A typical experimental result is shown in Fig.~\ref{fig:tomo_demo}, which corresponds to $m_z=1.4 t_0$, $t_{so}=0.2 t_0$, $(k_x,k_y,k_z)=(0.1,0.6,0.1)$, quenching $\gamma_0$ and measuring $\gamma_0$. The result correspond to a time-averaged spin polarization $ \overline{\left < \gamma_0 \right >}_0 $ of $ 0.423 \pm 0.056 $, and theory value is $0.460$.

\section{Data processing method}

\subsection{Reconstruction of the BIS}

\begin{figure}[t]
	\centering
	%	\fbox
	{\includegraphics[clip, trim=0cm 5cm 10cm 0cm, width=1.0\columnwidth]
		{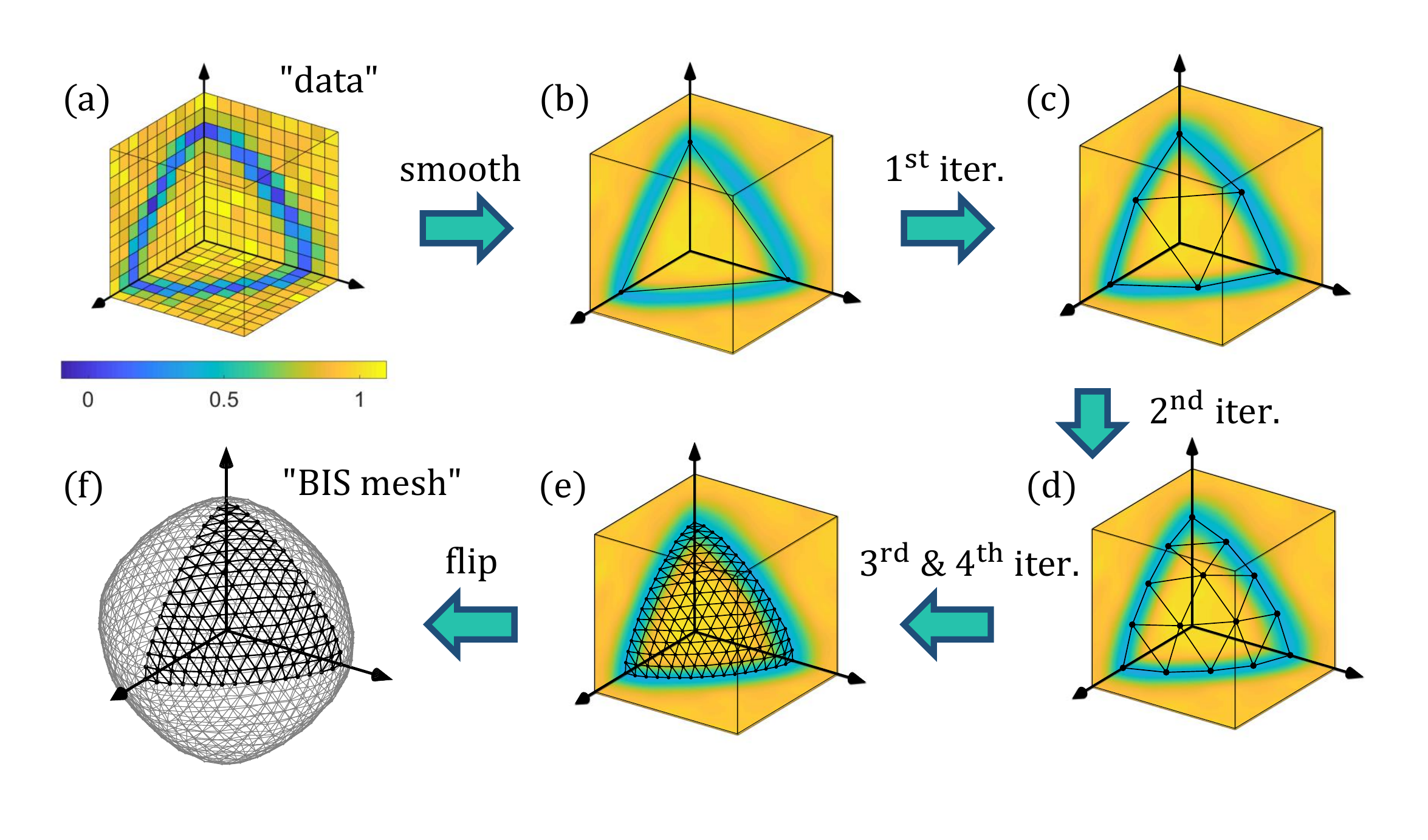}}
	\caption{
		\textbf{Reconstruction of BIS.}
		(a) The raw experimental results of $ \overline{\left < \gamma_0 \right >}_0 $ within the first octant. Note that, for clarity, only the $k_x=0$, $k_y=0$ and $k_z=0$ planes are plotted.
		(b) The smoothed and interpolated $ \overline{\left < \gamma_0 \right >}_0 $ field and the initial triangular mesh.
		(c,d,e) 2nd, 3rd and 5th iteration of triangular mesh.
		(f) Flip the mesh in (e) and combine them to form the full BIS.
	}
	\label{fig:bis}
\end{figure}

To obtain the BIS, we quench along $\gamma_0$ and measure $\overline{ \left< \gamma_0 \right> }$ at different $\textbf k$. Since the Hamiltonian is symmetric under $k_x$-, $k_y$-, and $k_z$-reflections, the $\overline{ \left< \gamma_0( \textbf{k} ) \right> }_0$ result in the first octant of the Brillouin zone is sufficient to reconstruct the BIS. We measure $\overline{ \left< \gamma_0( \textbf{k} ) \right> }_0$ in a mesh grid with a $0.1\pi$ step size for $k_x$, $k_y$ and $k_z$, which is sufficient to reconstruct the BIS well in our interested case.  As shown in Fig.~\ref{fig:bis}, the whole reconstruction process is based on data smoothing and iteratively interpolating a triangular mesh. Specifically, we first define an initial triangle, as a coarse representation of BIS, by finding the minimum of the smoothed $\overline{ \left< \gamma_0 \right> }_0$ field along all three axes. For each edge of the old mesh, we locate its center, and find the minimum of the $\overline{ \left< \gamma_0 \right> }_0$ field along the norm line of the old face at that location to define a new vertex. Combining the new vertices with the old ones, we can obtain a refined mesh and describe the BIS more accurately. With repeating of this process, we can reconstruct the BIS mesh in the first octant form the measurement result to any demanding accuracy. Finally, we flip the BIS mesh to other octants and combine them to obtain the full BIS mesh in the Brillouin zone.

\subsection{Measurement of g field winding number}

\begin{figure}[t]
	\centering
%	\fbox
	{\includegraphics[clip, trim=0cm 7cm 12cm 0cm, width=1.0\columnwidth]
		{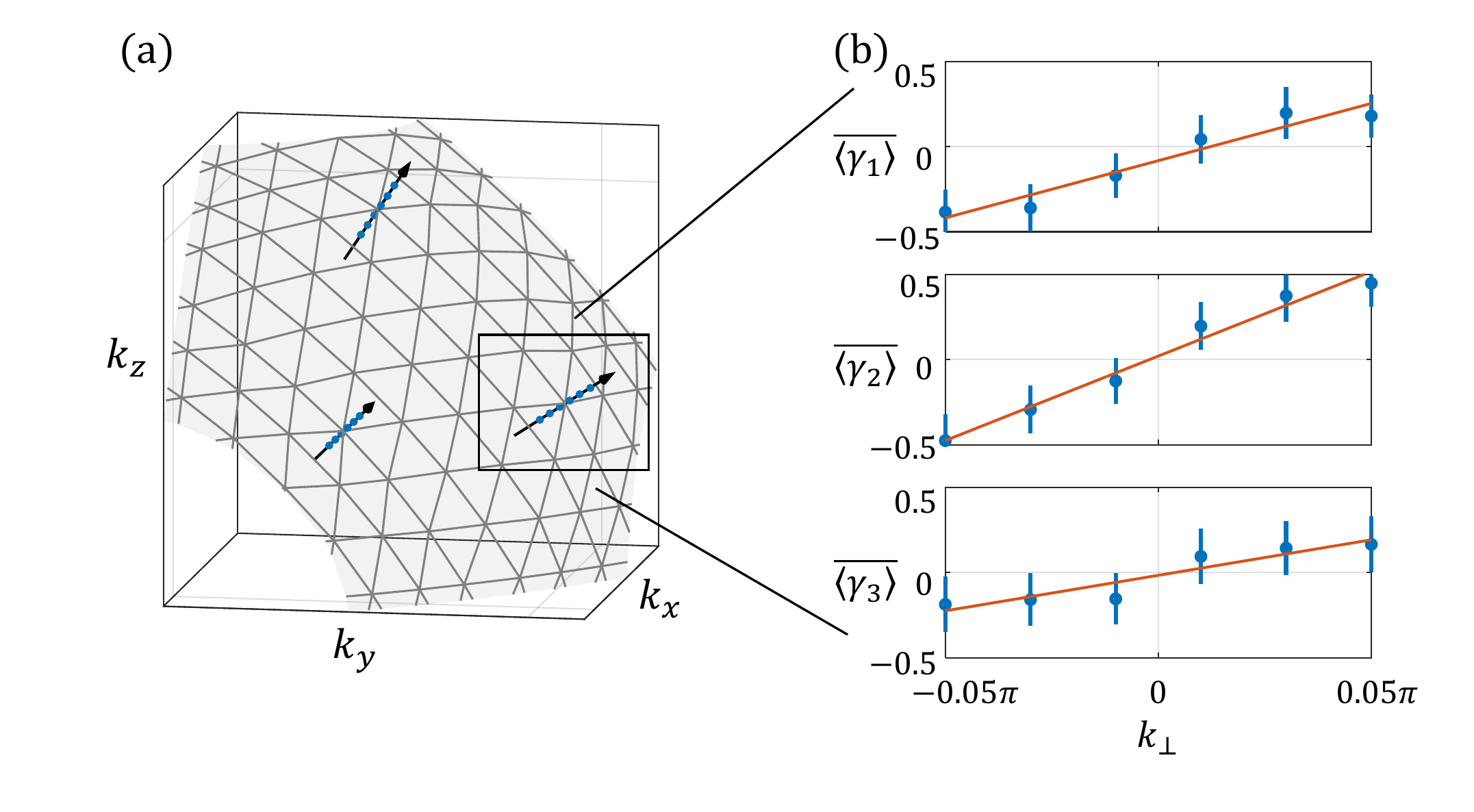}}
	\caption{
		\textbf{The measurement of g field.}
		(a) A closeup of three positions that $\textbf g$ field is measured. The arrow marks the norm direction of the BIS, along which 6 points are measured to fit for the slop.
		(b) $ \overline{\left < \gamma_{1,2,3} \right >}_0 $ measurement results along $\textbf k_\perp$, the error bars represent 3 standard deviation. The lines represent the slope fitted from the results.
	}
	\label{fig:gvec}
\end{figure}

Based on the previously obtained BIS mesh, we measure the emergent dynamical spin-texture field, i.e., the $\textbf g( \textbf k )$ field, of which the components are defined as
\begin{equation}
	g_i(\textbf k) = \frac{1}{\mathcal N_k} \partial_{\textbf k_\perp} \overline{ \left< \gamma_i( \textbf{k} ) \right> }.
\end{equation}
We measure the $\textbf g( \textbf k )$ field by sampling 6	 points across the BIS, along the norm direction, with a step size of $0.02\pi$, and measuring the time-averaged spin polarization $\overline{ \left< \gamma_{1,2,3} \right> }_0$. The slopes fitted from the results, after normalization, give the $\textbf g( \textbf k )$ field. A typical experimental result is depicted in Fig.~\ref{fig:gvec}.

The integral calculating the winding number of the $\textbf g( \textbf k )$ field is discretized as a summation over all the triangular meshes, i.e.,
\begin{align}
\mathcal{W}  = \frac{1}{8\pi}\int_{\rm BIS}\mathrm{d}^{2}\mathbf{k}\,\mathbf{g}\cdot(\nabla\mathbf{g}\times\nabla\mathbf{g}) = \frac{1}{4\pi} \sum_i S_i,
\end{align}
where $i$ is the label of triangular element and $S_i$ is the solid angle formed by the three $\textbf g$ vectors on the vertices of the $i$-th triangular element.
The solid angle is calculated by the sum of the three internal sphere angles subtracted by $\pi $.

For the symmetry-breaking case, the $g_4$ term is approximated by a constant within the same triangular element. From Eq.~\ref{eq:winding} we have
\begin{equation}
\begin{aligned}
\mathcal W_{\rm SB} & = \frac{1}{\pi^2}  \sum_{i} \int_{\mathcal S_i} \mathrm{d} S^2 \int_{0}^{\phi_{3,i}} \mathrm{d} \phi_3 \sin^2 \phi_3\\
& =   \frac{1}{\pi^2} \sum_{i} S_i \frac{\phi_{3,i} - \sin \phi_{3,i} \cos \phi_{3,i}}{2} .
\end{aligned}
\end{equation}
where $\mathrm{d} S^2$ is the area element of the 2-sphere, and the integral is taken within each triangular mesh, giving the solid angle formed by $(g_1,g_2,g_3)$. The $\mathcal W_{\rm SB}$ is calculated in the same way as $\mathcal W$, with each element multiplied by a factor depending on $g_4$.

\subsection{Error analysis}

The dominant error in our experiment comes from the shot noise in the the optical readout, which yields a normal distribution of the photon counts with a mean of $N$ and a standard deviation of $\sqrt{N}$, where $N$ is between 1000 to 2000 in the experiment for a fixed 10,000 repetitions of each sequence. This random distribution then introduces an uncertainty in obtaining the dynamical spin-texture field.

To estimate the error associated with a quantity, e.g. the winding number, we adopt the Monte Carlo method. First, we randomly generate photon counts of the same distribution with the measurements. Then we feed the generated counts to the algorithm for calculating the winding number. This process is repeated sufficient times, and we take the standard deviation of the results as the error of the winding number.

\section{Results with spin dephasing}
\label{sec:deco}

\begin{figure}[t]
	\centering
%	\fbox
	{\includegraphics[clip, trim=0cm 2cm 9cm 0cm, width=1\columnwidth]
		{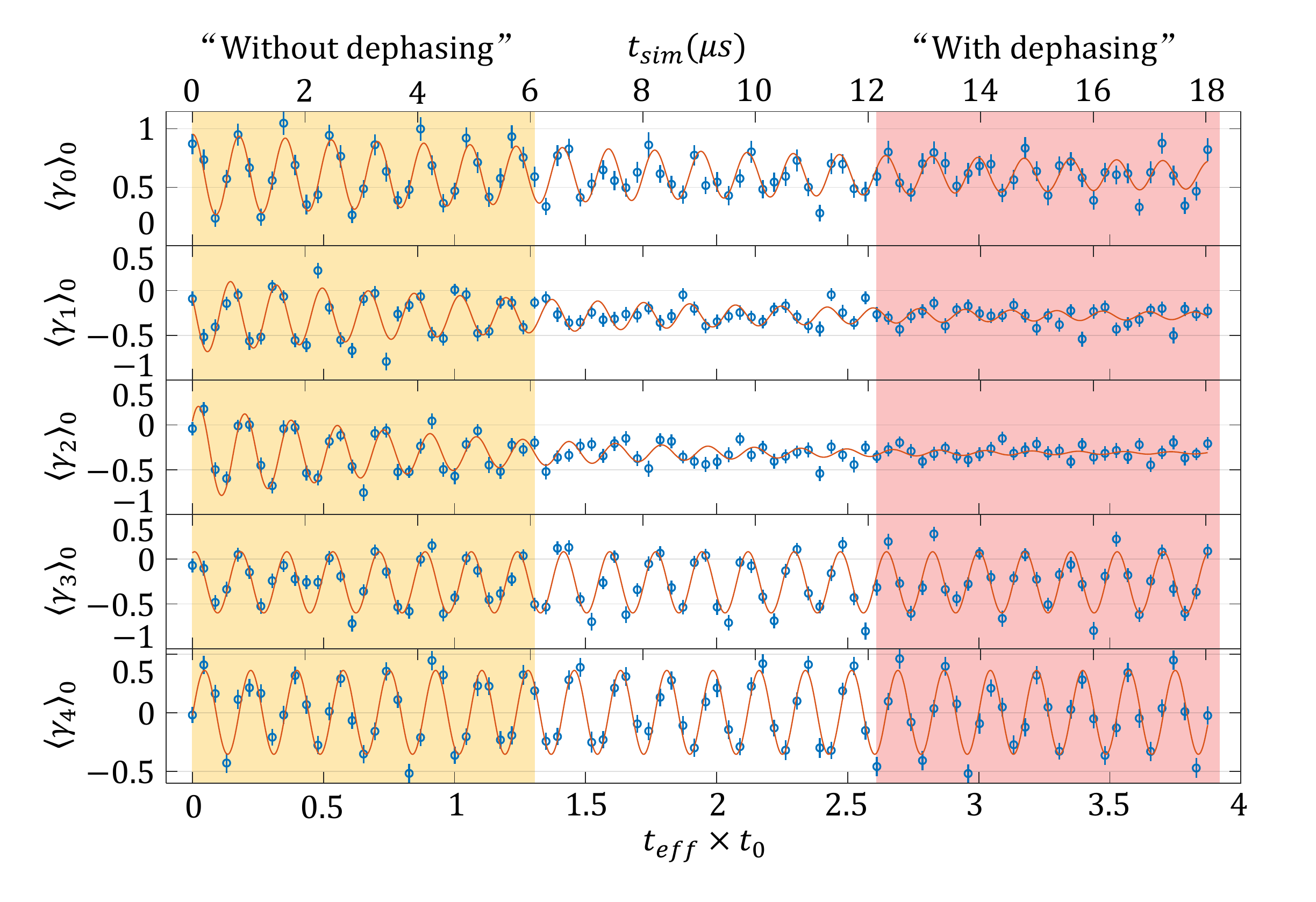}}
	\caption{
		\textbf{Dephasing in the evolution.}
		The results shown correspond to $m_z=1.4 t_0$, $t_{so}=t_0$, $ k_x = k_y = k_z = -0.6 \pi $, and deep quench along $\gamma_0$.
		The yellow (red) area denotes the time range from $0$ to $t_{\rm max}$ (from $2 t_{\rm max}$ to $3 t_{\rm max}$).
	}
	\label{fig:deco_demo}
\end{figure}

\begin{figure}[t]
	\centering
%	\fbox
	{\includegraphics[clip, trim=0cm 4cm 17cm 0cm, width=0.8\columnwidth]
		{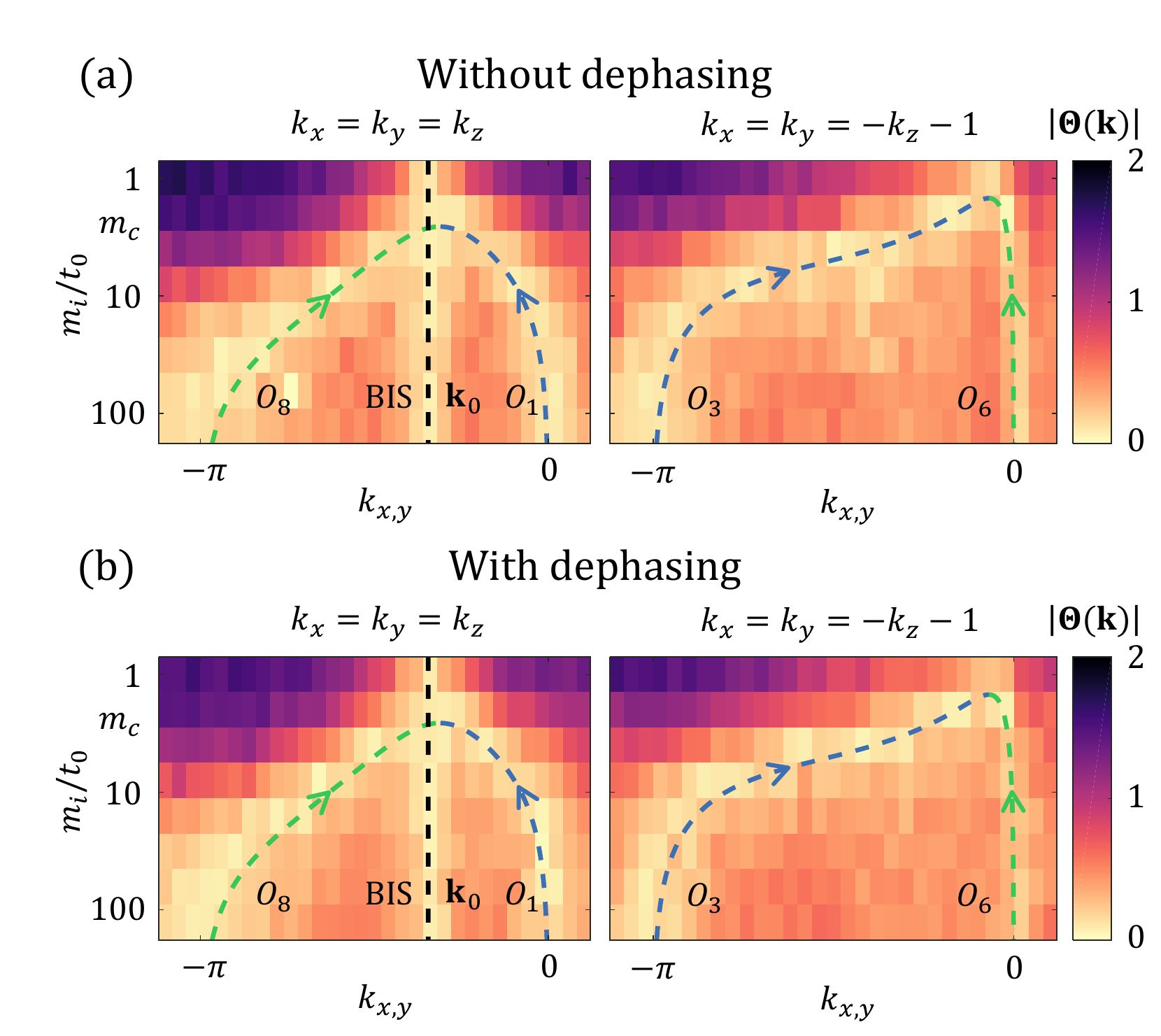}}
	\caption{
		\textbf{Measurement of charge movement with dephasing.}
		 Repeat the measurement in main text Fig. 4(b), with the time range chosen from $2 t_{\rm max}$ to $3 t_{\rm max}$ to introduce dephasing. The results are almost identical, showing the robustness of the dynamical characterization methods against dephasing.
	}
	\label{fig:deco}
\end{figure}

We investigate the effect of dephasing in the simulation by extending the evolution time. All the previous time-averaged spin polarization measurements are averaged with evolution time $ t $ from $0$ to $t_{\rm max}$ as determined by Eq.~\ref{eq:t_eff_max}), while the results with dephasing are averaged with $ t $ from $2 t_{\rm max}$ to $3 t_{\rm max}$.
As an example, we choose $m_z=1.4 t_0$, $t_{so}=t_0$, $ k_x = k_y = k_z = -0.6 \pi $, and deep quench along $\gamma_0$. $ \left < \gamma_{0,1,2,3,4} \right >_0 $ are measured for a series of evolution time, from $0$ to $3 t_{\rm max}$. The experimental results are shown in Fig.~\ref{fig:deco_demo}, with yellow (red) area denoting the time range from $0$ to $t_{\rm max}$ (from $2 t_{\rm max}$ to $3 t_{\rm max}$). One can easily see that although the amplitude of the oscillation damps, the average value maintains the same, which means that averaging over time range with or without dephasing will give the same result. The  $ \left < \gamma_{1,2} \right >_0 $ results decay fast, while the  $ \left < \gamma_{3,4} \right >_0 $ results decay only negligibly. This is due to the fact that $ \gamma_{1,2}  $ correspond to $\sigma_{x,y}$, which depend on the short electron spin coherence time, while $  \gamma_{3,4} $ depend on the nuclear spin, of which the coherence time is much longer than this time scale.

We further demonstrate this robustness against dephasing by repeating the measurement in main text Fig. 4(b), with time range chosen from $2 t_{\rm max}$ to $3 t_{\rm max}$. The comparison of results with and without dephasing is shown in Fig.~\ref{fig:deco}.

\end{document}